\documentclass{article}
\usepackage{amssymb}
\usepackage{amsmath}
\usepackage{amsfonts}
\usepackage{graphicx}

\newtheorem{theorem}{Theorem}[section]

\newtheorem{claim}[theorem]{Claim}

\newtheorem{definition}[theorem]{Definition}

\newtheorem{remark}[theorem]{Remark}

\input{tcilatex}

\begin{document}

\title{Global geometry of planary 3-body motions}
\author{Wu-Yi Hsiang \\
Department of Mathematics\\
University of California, Berkeley \and Eldar Straume \\
Department of Mathematical Sciences\\
Norwegian University of Science and \\
Technology, Trondheim, Norway}
\maketitle
\tableofcontents

\section{Introduction}

In a recent paper (cf. \cite{HS-1}) the authors have investigated 3-body
motions with vanishing angular momentum, in the framework of equivariant
Riemannian geometry and by resuming the basic approach dating back to
Jacobi's geometrization of Lagrange's least action principle, in the setting
of kinematic geometry of 3-body systems. A geometric reduction method was
described which reduces the study of trajectories of 3-body motions, first
from the level of the configuration space to the level of the moduli space
of congruence classes of m-triangles, and then

\begin{itemize}
\item further reduces the moduli curves to that of their shape curves on the
2-sphere $S^{2}$. Namely, a trajectory of 3-body motions is completely
determined, up to global congruence, by its shape curve which only records
the changing of shape (i.e. similarity class).

\item Moreover, the unique parametrization theorem further proves that the
trajectory is already determined by the geometric (i.e. non-parametrized)
shape curve.

\item Another remarkable property of the above shape curves is expressed by
the monotonicity theorem, concerning their (mass modified) latitude function
on the sphere $S^{2}.$
\end{itemize}

The monotonicity theorem is definitely only valid in the case of zero
angular momentum, but with this paper we shall extend the first two of the
above three stated results to the more general case of planary motions. We
start with a description of the basic setting from \cite{HS-1} in the two
subsections below, and a summary of the major results is presented in
Section 1.3.

In Section 2 we work out the kinematic and dynamical metric on the moduli
space, together with the associated differential equations. Finally, in
Section 3 we establish the remaining results needed for the proofs of the
two main theorems stated in Section 1.3.

\subsection{The basic kinematic quantities and the potential function}

The classical 3-body problem in celestial mechanics studies the local and
global geometry of the trajectories of a 3-body system, namely the motion of
three point masses (bodies) of mass $m_{i}>0$, say normalized to $\sum
m_{i}=1$, under the influence of the mutual gravitational forces. This
system constitutes a conservative mechanical system with the Newton's
potential function
\begin{equation}
U=\sum_{i<j}\frac{m_{i}m_{j}}{r_{ij}},\text{ \ \ \ }r_{ij}=\left\vert
\mathbf{a}_{i}-\mathbf{a}_{j}\right\vert \text{\ }  \label{1.1}
\end{equation}
and potential energy $-U$. We introduce the vector $\delta=(\mathbf{a}_{1},%
\mathbf{a}_{2},\mathbf{a}_{3})$, called an \emph{m-triangle}, which records
the position of the system in an inertial frame with the origin at the
center of mass, and hence $\sum m_{i}\mathbf{a}_{i}=0$.

A \emph{trajectory} is a time parametrized curve $\gamma(t)$ representing a
motion of the 3-body system, locally characterized by Newton's equation%
\begin{equation}
\frac{d^{2}}{dt^{2}}\gamma=\nabla U(\gamma)=(\frac{1}{m_{1}}\frac{\partial U%
}{\partial\mathbf{a}_{1}},\frac{1}{m_{2}}\frac{\partial U}{\partial \mathbf{a%
}_{2}},\frac{1}{m_{3}}\frac{\partial U}{\partial\mathbf{a}_{3}}\text{\ })%
\text{\ }  \label{1.2}
\end{equation}
However, the trajectories can also be characterized globally as solutions of
a suitable boundary value problem, characterized as extremals of an
appropriate least action principle, such as the two principles due to
Lagrange and Hamilton.

Let us also recall the basic kinematic quantities which are the (polar)
moment of inertia, kinetic energy and angular momentum, respectively defined
by
\begin{equation}
I=\sum m_{i}\left\vert \mathbf{a}_{i}\right\vert ^{2}\text{, \ }T=\frac{1}{2}%
\sum m_{i}\left\vert \mathbf{\dot{a}}_{i}\right\vert ^{2}\text{, \ \ \ }%
\mathbf{\Omega}=\sum m_{i}(\mathbf{a}_{i}\times\mathbf{\dot{a}}_{i})
\label{1.3}
\end{equation}
The dynamics of the 3-body problem is largely expressed by their
interactions with the potential function $U$ \ via the equation (\ref{1.2}),
and for example, the invariance of the total energy
\begin{equation}
h=T-U  \label{h}
\end{equation}
is a simple consequence of (\ref{1.2}) and the definition of $T$. On the
other hand, whereas the invariance of the vector $\mathbf{\Omega}$ follows
from the rotational symmetry of $U$, in this article we shall exploit the
consequences of the additional homogeneity property of $U$.

\subsection{Reduction to the moduli space and the shape space}

In this article we shall only be concerned with planary three-body motions,
namely the m-triangles $\delta$ are confined to a fixed plane $\mathbb{R}%
^{2} $ and hence belong to the \emph{configuration space}
\begin{equation}
M\simeq\mathbb{R}^{4}:\sum\limits_{i=1}^{3}m_{i}\mathbf{a}_{i}=0\text{, \ }%
\mathbf{a}_{i}\in\mathbb{R}^{2}  \label{M}
\end{equation}
With the inner product of m-triangles defined by

\begin{equation}
\delta\cdot\delta^{\prime}=\sum m_{i}\mathbf{a}_{i}\cdot\mathbf{b}_{i}
\label{metric}
\end{equation}
$M$ is given the \emph{kinematic metric}, namely the metric such that the
right side of Newton's equation (\ref{1.2}) is the gradient of $U$. Then the
squared norm is the moment of inertia, $I=I(\delta)=\left\vert \delta
\right\vert ^{2}$, and hence the \emph{hyperradius }$\rho=\sqrt{I}$\ is the
natural \emph{scaling function }which also measures the distance from the
origin.\emph{\ }

The rotation group $SO(2)$ acts naturally, and by orthogonal
transformations, on $M$ by rotating m-triangles, and the orbit spaces of $M$
(resp. its unit sphere $M^{1}\simeq S^{3}$) are the (congruence) \emph{%
moduli space }$\bar{M}$ (resp. the \emph{shape space} $M^{\ast}$), namely
\begin{equation}
\bar{M}=M/SO(2)\text{, \ }M^{\ast}=M^{1}/SO(2)\text{\ }  \label{orbitspace}
\end{equation}
The points in $\bar{M}$ represent \emph{congruence classes} $\bar{\delta}$
of m-triangles, and points in $M^{\ast}$ represent the \emph{shapes} (or
similarity classes) $\delta^{\ast}$ of non-zero m-triangles.

Geometrically,the above orbit space construction and orbit map $M\rightarrow$
$\bar{M}$ is, in fact, just the Hopf map construction, whose restriction $%
S^{3}\rightarrow S^{2}$ is the classical Hopf fibration. Namely, the spaces
fit into the following diagram
\begin{equation}
\begin{array}{ccc}
M\simeq & \mathbb{R}^{4} & \longrightarrow\bar{M} \\
\cup & \cup & \cup \\
M^{1}\simeq & S^{3} & \rightarrow M^{\ast}\simeq S^{2}(1/2)%
\end{array}
\label{diagram}
\end{equation}
where $M\simeq\mathbb{R}^{4}$ is a chosen $SO(2)$-equivariant isometry (i.e.
choice of Jacobi vectors). In particular, $\bar{M}$ $\approx\mathbb{R}^{3}$
is a cone over $M^{\ast}$ and there is the radial projection $\bar{M}%
-\left\{ O\right\} \rightarrow M^{\ast}$ which "reduces" a non-zero
congruence class $\bar{\delta}$ to its shape $\delta^{\ast}$. Note, however,
the representation of the various shapes of m-triangles on a fixed model
sphere $S^{2}$ depends on the mass distribution $\left\{ m_{i}\right\} $,
via the mass dependence of the Jacobi vectors.

Briefly, in this article we shall analyze the two-step reduction%
\begin{equation}
M\rightarrow\bar{M}\text{, \ }\bar{M}-\left\{ O\right\} \text{\ }\rightarrow
M^{\ast}\text{, \ \ \ }\gamma(t)\rightarrow\bar{\gamma}(t)\text{\ }%
\rightarrow\gamma^{\ast}(t)  \label{redu}
\end{equation}
by which a trajectory $\gamma(t)$ of a planary 3-body motion is projected to
its moduli curve $\bar{\gamma}(t)$ and further to its shape curve $%
\gamma^{\ast}(t)$ on a 2-sphere. In Section 2.1 we shall put the above
reduction and the spaces involved in the framework of Riemannian geometry,
and moreover, explain how Jacobi's geometrization idea can be reduced and
extended to the level of $\bar{M}$.

\subsection{A summary of the main results}

The Hopf map construction (\ref{diagram}) makes it convenient to use a
Euclidean model, $\bar{M}$ $=\mathbb{R}^{3}$, for the moduli space and with
the unit sphere $S^{2}(1)$ as the shape space $M^{\ast}.$ In this way one
can express all kinematic quantities and dynamical equations in terms of
spherical geometry and spherical coordinates, and hence take the full
advantage of the cone structure of $\bar{M}$ over $M^{\ast}$.

One can start from Newton's equation (\ref{1.2}) for planary m-triangles
\begin{equation}
\frac{d^{2}}{dt^{2}}\gamma=\nabla U(\gamma),\text{ \ \ }\gamma(t)\in M\text{%
\ }  \label{S3}
\end{equation}
with any $SO(2)$-invariant potential function $U$, and hence it is a
function on $\bar{M}$. The additional crucial property of $U$ that we have
exploited is its \emph{homogeneity}, namely it is of type
\begin{equation}
U=\frac{U^{\ast}(\varphi,\theta)}{\rho^{e}}  \label{S2}
\end{equation}
where $U^{\ast}$ denotes the restriction of $U$ to the sphere $M^{\ast}$.
The Newtonian case $e=1$ is certainly the most important one, but the proofs
are essentially the same for other (integral) values of $e>0$.

Consider trajectories $\gamma(t)$ of (\ref{S3}) for a given energy-momentum
level $(h,\omega)$, $\omega=\left\vert \Omega\right\vert $, and for
spherical coordinates $(\rho,\varphi,\theta)$ in $\bar{M}$, let the curves
\begin{equation}
\bar{\gamma}(t)=(\rho(t),\gamma^{\ast}(t)),\text{ \ \ }\gamma^{\ast
}(t)=(\varphi(t),\theta(t))  \label{S1}
\end{equation}
be the associated moduli and shape curve, respectively. In Section 2.2 we
show the \emph{reduced Newton's equation} in $\bar{M}$ can be presented as
the pair
\begin{equation}
\ddot{I}=2(U+2h)+2(1-e)U,\text{ \ \ \ }\ddot{\gamma}^{\ast}+P\dot{\gamma }%
^{\ast}+Q\nabla U^{\ast}=0\text{\ \ \ }  \label{S5}
\end{equation}
where the first equation in (\ref{S5}) is simply the Lagrange-Jacobi
equation, and $\ddot{\gamma}^{\ast}$ is the covariant acceleration of $%
\gamma^{\ast}$ as a spherical curve. \ Moreover, the energy integral (\ref{h}%
) is the following first order equation in $\bar{M}$%
\begin{equation}
\frac{1}{2}\dot{\rho}^{2}+\frac{\rho^{2}}{8}v^{2}+\frac{\omega^{2}}{2\rho^{2}%
}-\frac{U^{\ast}}{\rho^{e}}-h=0  \label{S7}
\end{equation}
where $v=\left\vert \dot{\gamma}^{\ast}\right\vert =$ $\sqrt{\dot{\varphi}%
^{2}+(\sin^{2}\varphi)\dot{\theta}^{2}}$ is the speed of the shape curve. In
fact, combined with (\ref{S7}) any of the three scalar equations in (\ref{S5}%
) can be derived from the other ones. The equations of (\ref{S5}) are
presented in the coordinates $(\rho,\varphi,\theta)$ in Section 2.2.

On the other hand, let $K^{\ast}$ be the geodesic curvature of $%
\gamma^{\ast} $ and $U_{\mathbf{\nu}}^{\ast}$ the directional derivative of $%
U^{\ast}$ normal to $\gamma^{\ast}$.Then there is the formula
\begin{equation}
\rho^{2+e}=\frac{4}{v^{2}}\mathfrak{S,}\text{ \ \ where }\mathfrak{S=}\frac{%
U_{\mathbf{\nu}}^{\ast}}{K^{\ast}}\text{ \ \ \ (cf. }(\ref{S8}))  \label{S4}
\end{equation}
which separates the radial variable $\rho$ from the spherical ones. Using (%
\ref{S4}), the dependence on $\rho$ in the coefficient functions $P,Q$ in (%
\ref{S5}) can be eliminated, namely
\begin{equation}
P=2\frac{\dot{\rho}}{\rho}=\frac{2}{2+e}\frac{\mathfrak{\dot{S}}}{\mathfrak{S%
}}-\frac{4}{2+e}\frac{\dot{v}}{v},\text{ \ \ \ }Q=-\frac{4}{\rho^{2+e}}=-%
\frac{v^{2}}{\mathfrak{S}},  \label{S6}
\end{equation}
which yields a third order equation for $\gamma^{\ast}$ which is, in fact,
independent of $(h,\omega).$

Clearly, the above function $\mathfrak{S}$ depends only on the relative
geometry between $\gamma^{\ast}$ and the gradient of $U^{\ast}$. However, we
regard $\mathfrak{S}$ as undefined if $\gamma^{\ast}$ is a geodesic arc (and
hence lies on a gradient line), and any such solution of (\ref{S5}) is
called \emph{exceptional}$.$ Now, assuming (for simplicity) that the shape
curve is not of exceptional type, our main results can be formulated neatly
as the following two theorems :

\begin{theorem}
\label{A}For a given total energy and nonzero angular momentum, a planary
three-body motion is completely determined up to congruence by its time
parametrized shape curve (which only records the changing of shape).
\end{theorem}

\begin{theorem}
\label{B}The time parametrization is uniquely determined by the relative
geometry between the oriented geometric (i.e. non-parametrized) shape curve
and the gradient vector field of $U^{\ast}$.
\end{theorem}

\begin{remark}
(i) In the case of non-zero total energy, Theorem \ref{A} remains unchanged
in the case of zero angular momentum, whereas the motion is determined up to
congruence and scaling in the case of ($h,\omega)=(0,0)$. We refer to \cite%
{HS-1}, Section 4.2.

(ii) Uniqueness of time parametrization means, of course, modulo time
translation, or modulo an affine time transformation when $(h,\omega)=(0,0)$.

(iii) The proofs of the above theorems are the same for any homogeneous
potential function of type (\ref{S2}), with $e>0,$ and $e$ integral in
Theorem \ref{B}. We choose the most important case, $e=1$, in Section 3.2
and work out the crucial details. However, the formulae are even simpler in
the case $e=2$.
\end{remark}

\section{Riemannian geometry and reduction to the moduli space}

\subsection{Riemannian structures on the moduli space $\bar{M}$}

In his famous lectures \cite{Jac}, Jacobi introduced the concept of a \emph{%
kinematic metric }$ds^{2}$ on the configuration space $M$ of a mechanical
system with kinetic energy $T.$ For example, in the case of an n-body system
with total mass $\sum m_{i}=1,$\emph{\ }
\begin{equation}
ds^{2}=2Tdt^{2}=\sum\limits_{i}m_{i}(dx_{i}^{2}+dy_{i}^{2}+dz_{i}^{2})
\label{2.1}
\end{equation}
which is clearly equivalent to the definition (\ref{metric}). Now, for a
system with potential energy $-U$ and a fixed total energy $h$, set
\begin{align}
M_{h} & =\left\{ p\in M;h+U(p)\geq0\right\}  \label{2.3} \\
ds_{h}^{2} & =(h+U)ds^{2}  \notag
\end{align}
where $ds_{h}^{2}$ is called the \emph{dynamical metric} on $M_{h}$. By
writing
\begin{equation*}
ds_{h}=\sqrt{h+U}ds=\sqrt{T}ds=\sqrt{2}Tdt
\end{equation*}
Jacobi transformed Lagrange's action integral (on the left side of (\ref{2.4}%
)) into an arc-length integral, namely
\begin{equation}
J(\gamma)=\int_{\gamma}Tdt=\frac{1}{\sqrt{2}}\int_{\gamma}ds_{h}  \label{2.4}
\end{equation}
and hence the least action principle becomes the following simple geometric
statement :
\begin{align}
& \text{" Trajectories with total energy }h\text{ are exactly those \emph{%
geodesic curves}}  \label{state1} \\
\text{ } & \text{in the }\text{space }M_{h}\text{ with the dynamical metric }%
ds_{h}^{2\text{ }}\text{"}  \notag
\end{align}

Nowadays, the metric spaces $(M,ds^{2}),(M_{h},ds_{h}^{2\text{ }})$ are
called $\emph{Riemannian}$ $\emph{manifolds}$, and the dynamical metric is a
conformal modification of the kinematic metric by the scaling function $%
(U+h) $. As exemplified by (\ref{2.1}), a Riemannian metric on a manifold $N$
amounts to the choice of a kinetic energy function on the tangent bundle, $%
T:TN\rightarrow\mathbb{R}$, which is a positive definite quadratic form on
each tangent plane $T_{p}N$. This allows us to define the arc-length
function $u(t)$ and the kinetic energy along a given time parametrized curve
$\Gamma(t)$ by%
\begin{equation}
T(t)=\frac{1}{2}(\frac{du}{dt})^{2}=\frac{1}{2}\left\vert \frac{d\Gamma}{dt}%
\right\vert ^{2}  \label{2.7}
\end{equation}

Now, let us determine the appropriate kinetic energy $\bar{T}$ on the moduli
space $\bar{M}$ and hence also define its \emph{kinematic metric} using the
recipe (\ref{2.7}). At the same time, referring to the diagram (\ref{diagram}%
) and the Hopf map, let us also introduce the \emph{orbital distance metric }%
$d\bar{s}^{2}$ on $\bar{M}$ as an $SO(2)$-orbit space. Then $(\bar{M},d\bar{s%
}^{2})$ inherits the structure of a Riemannian cone over the shape space $%
(M^{\ast},d\sigma^{2})$, namely%
\begin{equation}
d\bar{s}^{2}=d\rho^{2}+\rho^{2}d\sigma^{2}\text{, \ \ }d\sigma^{2}=d\bar {s}%
^{2}|_{M^{\ast}}  \label{2.8}
\end{equation}
Moreover, it is well known that the Hopf fibration in the above Riemannian
setting is $S^{3}(1)\rightarrow S^{2}(1/2)$, and consequently
\begin{equation}
(M^{\ast},d\sigma^{2})\simeq S^{2}(1/2)  \label{2.9}
\end{equation}
is also the round sphere of radius $1/2$.

Consider a curve $\gamma(t)$ in $M$ and its orthogonal velocity
decomposition $\dot{\gamma}=\dot{\gamma}^{h}+\dot{\gamma}^{\omega}$ and
corresponding splitting of kinetric energy%
\begin{equation}
T=T^{h}+T^{\omega},  \label{2.10}
\end{equation}
where $\dot{\gamma}^{\omega}$ is tangential to the $SO(2)$-orbit, and hence $%
T^{\omega}$ is the kinetic energy due to purely rotational motion of
m-triangles. By definition of the metric $d\bar{s}^{2}$, the orbit map $%
M\rightarrow$ $\bar{M}$ is a Riemannian submersion and hence maps the
"horizontal" component $\dot{\gamma}^{h}$ isometrically to the velocity
vector of $\bar{\gamma}$. This shows $T^{h}=\bar{T}$ is also the kinetic
energy at the level of $\bar{M}$, that is, the kinematic metric coincides
with the orbital distance metric, and by (\ref{2.8}), (\ref{2.9}) and (\ref%
{2.10}) the latter can be finally expressed as
\begin{align}
d\bar{s}^{2} & =2\bar{T}dt^{2}=2(T-T^{\omega})dt^{2}=2(T-\frac{\omega^{2}}{%
2\rho^{2}})dt^{2}  \label{2.11} \\
& =d\rho^{2}+\rho^{2}d\sigma^{2}=d\rho^{2}+\frac{\rho^{2}}{4}(d\varphi
^{2}+\sin^{2}\varphi d\theta^{2})  \notag
\end{align}

\begin{remark}
The expression $d\varphi^{2}+\sin^{2}\varphi d\theta^{2}$ in the last line
of (\ref{2.11}) is the metric of the unit sphere $S^{2}(1)$ in terms of
spherical polar coordinates. In fact, the metric $d\bar{s}^{2}$ differs from
\ the Euclidean metric only by the factor $1/4$ in (\ref{2.11}), which makes
it singular at the origin. Moreover, $d\bar{s}^{2}$ is actually a conformal
modification of the Euclidean metric (cf. \cite{HS-1}, Section 2).
\end{remark}

Next, we turn to the construction of the dynamical metric on the moduli
space $\bar{M}$, which depends on $U$ and a given energy-momentum level $%
(h,\omega )$. Following the geometrization idea of Jacobi, we want the
geodesics of this metric to be the trajectories in $\bar{M}$, regarded as a
simple mechanical system with kinetic energy $\bar{T}$, potential energy $%
\bar{U}$, and conserved total energy $h=\bar{T}-\bar{U}$. Thus we introduce
the reduced potential function on $\bar{M}$
\begin{equation*}
\bar{U}=U-\frac{\omega^{2}}{2\rho^{2}}
\end{equation*}
and define the \emph{dynamical metric }%
\begin{equation}
d\bar{s}_{h,\omega}^{2}=\bar{T}d\bar{s}^{2}=(\bar{U}+h)d\bar{s}^{2}=(U+h-%
\frac{\omega^{2}}{2\rho^{2}})d\bar{s}^{2}  \label{2.12}
\end{equation}

Finally, it is not difficult to see that Lagrange's least action principle (%
\ref{2.4}) as well as Hamilton's least action principle using the Lagrange
function $L=T+U$, can be pushed down to the level of $\bar{M}$. This yields
the function $\bar{L}=\bar{T}+\bar{U}$, and for example, by following
Jacobi's geometrization idea applied to Lagrange's action integral in $\bar{M%
}$
\begin{equation*}
\bar{J}(\bar{\gamma})=\sqrt{2}\int_{\bar{\gamma}}\bar{T}dt=\sqrt{2}\int _{%
\bar{\gamma}}(\bar{U}+h)dt=\int_{\bar{\gamma}}\sqrt{\bar{U}+h}d\bar{s}=\int d%
\bar{s}_{h,\omega}\text{ ,}
\end{equation*}
we arrive at the following geometric statement similar to (\ref{state1}) :%
\begin{align}
& \text{"Curves in }\bar{M}\text{ representing trajectories in }M\text{ at a
given }  \notag \\
& \text{energy-momentum }\text{level }(h,\omega)\text{ are exactly those
\emph{geodesic}}  \label{state2} \\
& \text{\emph{\ curves} in }\bar{M}\text{ with the induced dynamical metric }%
d\bar{s}_{h,\omega}^{2}\text{."}  \notag
\end{align}

\subsection{The geodesic equations of the moduli space}

The moduli space $\bar{M}$ is, first of all, equipped with the kinematic
metric
\begin{equation}
d\bar{s}^{2}=2\bar{T}dt^{2}=2(T-\frac{\omega^{2}}{2\rho^{2}})dt^{2}=d\rho
^{2}+\frac{\rho^{2}}{4}(d\varphi^{2}+\sin^{2}\varphi d\theta^{2})
\label{3.10}
\end{equation}
and for each energy-momentum level $(h,\omega)$ there is the following
conformal modification of $d\bar{s}^{2}$
\begin{equation}
d\bar{s}_{(h,\omega)}^{2}=\bar{T}d\bar{s}^{2}=(\bar{U}+h)d\bar{s}^{2}
\label{3.11}
\end{equation}
The latter is the \emph{dynamical metric }which characterizes those moduli
curves $\bar{\gamma}(t)$ representing trajectories $\gamma(t)$ at the
specified level $(h,\omega).$ Namely, $\bar{\gamma}$ is a geodesic of the
Riemannian metric (\ref{3.11}), which in the spherical coordinates $%
(\rho,\varphi,\theta)$ expresses as
\begin{equation*}
d\bar{s}_{(h,\omega)}^{2}=(\frac{U^{\ast}(\varphi,\theta)}{\rho^{e}}+h-\frac{%
\omega^{2}}{2\rho^{2}})[d\rho^{2}+\frac{\rho^{2}}{4}(d\varphi
^{2}+\sin^{2}\varphi d\theta^{2})]
\end{equation*}

The standard procedure for the derivation of the geodesic equations, via the
calculation of the corresponding Christoffel symbols, yields the following
system of equations expressed with respect to time as the independent
variable :
\begin{align}
(i)\text{ \ }0 & =\ddot{\rho}+\frac{\dot{\rho}^{2}}{\rho}-\frac{1}{\rho }(%
\frac{2-e}{\rho^{e}}U^{\ast}+2h)  \notag \\
(ii)\text{ \ }0 & =\text{\ }\ddot{\varphi}+2\frac{\dot{\rho}}{\rho}\dot{%
\varphi}-\frac{1}{2}\sin(2\varphi)\dot{\theta}^{2}-\frac{4}{\rho^{2+e}}%
U_{\varphi}^{\ast}  \label{3.12} \\
(iii)\text{ \ \ }0 & =\ddot{\theta}+2\frac{\dot{\rho}}{\rho}\dot{\theta }%
+2\cot(\varphi)\dot{\varphi}\dot{\theta}-\frac{4}{\rho^{2+e}}\frac{1}{%
\sin^{2}\varphi}U_{\theta}^{\ast}  \notag
\end{align}
Note that equation (i), associated with the radial variable $\rho$ of $%
\bar
{M}$, as a cone over the sphere $M^{\ast}=S^{2}$, is simply the
Lagrange-Jacobi equation, cf. (\ref{S5}). Moreover, the second equation in (%
\ref{S5}) is merely a reformulation of \ equation (ii) and (iii), as
explained in \cite{HS-1}, Section 3.4.2. On the other hand, the dependence
on $\omega$ in the above equations is only implicit, but it appears in
equation (i) via substitution of the energy integral%
\begin{equation}
(iv)\text{ \ }h=\bar{T}-\bar{U}=\frac{1}{2}\dot{\rho}^{2}+\frac{\rho^{2}}{8}(%
\dot{\varphi}^{2}+\sin^{2}\varphi\dot{\theta}^{2})+\frac{\omega^{2}}{%
2\rho^{2}}-\frac{U^{\ast}}{\rho^{e}}  \label{3.13}
\end{equation}
once we have specified the value of $h$. Equation (iv) makes any of the
three equations of (\ref{3.12}) superfluous and may be replaced by (iv), as
the first step of integration, with $\omega$ appearing as an integration
constant.

Let us also describe another approach to derive the ODEs in (\ref{3.12}),
namely by regarding $\bar{M}$ as a simple conservative mechanical system
with the Lagrange function%
\begin{equation}
\bar{L}=\bar{T}+\bar{U}=T+U-\frac{\omega^{2}}{\rho^{2}}=\frac{1}{2}\dot{\rho
}^{2}+\frac{\rho^{2}}{8}(\dot{\varphi}^{2}+\sin^{2}\varphi\dot{\theta}^{2})-%
\frac{\omega^{2}}{2\rho^{2}}+\frac{U^{\ast}}{\rho^{e}}  \label{3.14}
\end{equation}
Then, straighforward calculations of the associated Lagranges's equations%
\begin{equation}
\frac{d}{dt}(\frac{\partial\bar{L}}{\partial\dot{\rho}})=\frac{\partial\bar {%
L}}{\partial\rho}\text{, \ \ }\frac{d}{dt}(\frac{\partial\bar{L}}{\partial%
\dot{\varphi}})=\frac{\partial\bar{L}}{\partial\varphi},\text{ \ }\frac{d}{dt%
}(\frac{\partial\bar{L}}{\partial\dot{\theta}})=\frac {\partial\bar{L}}{%
\partial\theta}  \label{3.15}
\end{equation}
yield the system (\ref{3.12}). Similar calculations are worked out in \cite%
{HS-1}, Section 3.2.

\section{The proofs of Theorem \protect\ref{A} \ and Theorem \protect\ref{B}}

\subsection{Separation of the scaling variable}

Let $\gamma^{\ast}(t)=(\varphi(t),\theta(t))$ be a given time parametrized
curve on the unit sphere $S^{2}$, and set $s=s(t)\geq0$ to be its arc-length
function. Then its unit tangent and positively oriented unit normal are,
respectively%
\begin{equation}
\mathbf{\tau}^{\ast}=\frac{d\gamma^{\ast}}{ds}=\frac{1}{v}(\dot{\varphi}%
\frac{\partial}{\partial\varphi}+\dot{\theta}\frac{\partial}{\partial\theta }%
)\text{, \ }\mathbf{\nu}^{\ast}=\frac{1}{v}(-\dot{\theta}\sin\varphi \frac{%
\partial}{\partial\varphi}+\dot{\varphi}\frac{1}{\sin\varphi}\frac{\partial}{%
\partial\theta})  \label{3.1}
\end{equation}
and its speed and scalar acceleration are, respectively%
\begin{equation}
v=\sqrt{\dot{\varphi}^{2}+\sin^{2}\varphi\dot{\theta}^{2}}\text{, \ }\dot {v}%
=\frac{d}{dt}v=\frac{1}{v}[\dot{\varphi}\ddot{\varphi}+(\sin\varphi
\cos\varphi)\dot{\varphi}\dot{\theta}^{2}+\sin^{2}(\varphi)\dot{\theta}\ddot{%
\theta}]  \label{3.2}
\end{equation}
One way to calculate the geodesic curvature function $K^{\ast}$ is to
express $\gamma^{\ast}$ in Euclidean coordinates as $\mathbf{x}%
(s)=(x(s),y(s),z(s))$ and use the formula%
\begin{equation*}
K^{\ast}(s)=\mathbf{x}(s)\times\mathbf{x}^{\prime}(s)\cdot\mathbf{x}%
^{\prime\prime}(s)
\end{equation*}
where differentiation is with respect to $s$. Then, by returning to
spherical coordinates
\begin{align}
K^{\ast} & =(\cos\varphi)\theta^{\prime}(1+\varphi^{\prime2})+\sin
\varphi(\varphi^{\prime}\theta^{\prime\prime}-\theta^{\prime}\varphi
^{\prime\prime})  \notag \\
& =\frac{1}{v^{3}}\left\{ (\cos\varphi)\dot{\theta}(v^{2}+\dot{\varphi}%
^{2})+\sin\varphi(\dot{\varphi}\ddot{\theta}-\dot{\theta}\ddot{\varphi }%
)\right\}  \label{3.3}
\end{align}

Next, let us eliminate the second order terms $\ddot{\varphi}$ and $\ddot{%
\theta}$ in the expression (\ref{3.3}), using equations (ii), (iii) of the
system (\ref{3.12}). This procedure yields
\begin{align*}
K^{\ast}v^{3} & =(\cos\varphi)\dot{\theta}(v^{2}+\dot{\varphi}%
^{2})+(\sin\varphi)\dot{\varphi}\left( -\frac{2\dot{\rho}}{\rho}\dot{\theta }%
-2(\cot\varphi)\dot{\varphi}\dot{\theta}+\frac{4}{\rho^{2+e}}\frac{1}{%
\sin^{2}\varphi}U_{\theta}^{\ast}\right) \\
& -(\sin\varphi)\dot{\theta}\left( -\frac{2\dot{\rho}}{\rho}\dot{\varphi }+%
\frac{1}{2}\sin(2\varphi)\dot{\theta}^{2}+\frac{4}{\rho^{2+e}}U_{\varphi
}^{\ast}\right) \\
& =\frac{4}{\rho^{2+e}}\left( \frac{\dot{\varphi}}{\sin\varphi}U_{\theta
}^{\ast}-\dot{\theta}\sin\varphi U_{\varphi}^{\ast}\right) =\frac{4v}{%
\rho^{2+e}}U_{\mathbf{\nu}}^{\ast}
\end{align*}
and consequently we arrive at the formula
\begin{equation}
\rho^{2+e}=\frac{4}{v^{2}}\frac{U_{\mathbf{\nu}}^{\ast}}{K^{\ast}}=\frac {4}{%
v^{2}}\mathfrak{S}  \label{S8}
\end{equation}
Note that the function $\mathfrak{S=S(\gamma}^{\ast})$, called the Siegel
function in \cite{HS-1}, depends only on the intrinsic geometry of the pair $%
(\mathfrak{\gamma}^{\ast},U^{\ast})$ on the sphere.

\subsection{ Intrinsic geometry of the shape curve and the gradient of $%
U^{\ast}$}

\ In the local analysis of the moduli and the shape curve, and their
interaction with the potential function $U^{\ast}$, we shall distinguish
between two types of variables or quantities associated with a given moduli
curve $\bar{\gamma}(t)$ $=(\rho(t)$ $,\gamma^{\ast}(t))$. On the one hand,
the \emph{intrinsic} quantities depend only on $\gamma^{\ast}$ as an
oriented geometric (i.e. unparametrized) curve and $U^{\ast}$ as a function
on $S^{2}$, and on the other hand, the \emph{variable} quantities are
defined along $\bar{\gamma}$ or $\gamma^{\ast}$, depending on the scaling
function $\rho$ in the moduli space $\bar{M}$ or the time parametrization of
the curves.

The basic intrinsic quantities are the gradient field $\nabla U^{\ast}$ (or
its tangential and normal derivatives $U_{\tau}^{\ast}$, $U_{\nu}^{\ast}$),
the orthonormal frame field $\left\{ \mathbf{\tau}^{\ast},\mathbf{\nu }%
\right\} $ along $\gamma^{\ast}$, and the geodesic curvature function $%
K^{\ast}$ of $\gamma^{\ast}$. In general, the linkage between $\gamma^{\ast}$
and $U^{\ast}$ is neatly encoded into the intrinsic function $\mathfrak{S}%
=U_{\nu}^{\ast}/K^{\ast}$, introduced in (\ref{S8}), so we shall assume $%
\gamma^{\ast}$ is not confined to a geodesic circle (in which case $%
\mathfrak{S}$ is undefined).

We choose a (generic) point $P_{0}$ on $\gamma^{\ast}$, and let $s$ be the
arc-length parameter of $\gamma^{\ast}$ in the positive direction starting
from $P_{0}.$Then the coefficients of the following power series expansions $%
\ $
\begin{align}
K^{\ast} & =K_{0}+K_{1}s+K_{2}s^{2}+...  \notag \\
U^{\ast} & =u_{0}+\bar{u}_{1}s+\bar{u}_{2}s^{2}+...  \notag \\
U_{\tau}^{\ast} & =\bar{u}_{1}+2\bar{u}_{2}s+3\bar{u}_{3}s^{2}+...
\label{3.16} \\
U_{\nu}^{\ast} & =\omega_{0}+\omega_{1}s+\omega_{2}s^{2}+...  \notag \\
\mathfrak{S} & =\mathfrak{S}_{0}+\mathfrak{S}_{1}s+\mathfrak{S}_{2}s^{2}+...
\notag
\end{align}
yield intrinsic quantities (or geometric data) localized at the point $%
P_{0}. $The coefficients $\mathfrak{S}_{n}$ are expressible as rational
functions of $K_{i}$ and $\omega_{i}$, and generally, let us say the \emph{%
order} of a coefficient in (\ref{3.16}) is the highest order of derivatives
of local coordinates in its expression. Thus, we say $\varphi_{0},\theta_{0}$
and $u_{0}$ are the intrinsic geometric data of order 0 at $P_{0}$, and for
example, $\bar{u}_{1},\omega_{0}$ and $\mathbf{\tau}^{\ast}|_{P_{0}}$ have
order $1$, and $\omega_{n},\bar{u}_{n+1}$ (resp. $K_{n},\mathfrak{S}_{n}$)
have order $n+1$ (resp. $n+2)$.

Let $(\varphi,\theta)$ denote spherical polar coordinates so that $P_{0}$ is
different from any of the "poles" $\varphi=0$ or $\pi$. We shall expand the
coordinate functions of $\bar{\gamma}(t)$, as well as $U^{\ast}$ and its
partial derivatives, as power series with respect to $t:$
\begin{align}
\rho & =\rho_{0}+\rho_{1}t+\rho_{2}t^{2}+\rho_{3}t^{3}+....  \notag \\
\varphi & =\varphi_{0}+\varphi_{1}t+\varphi_{2}t^{2}+\varphi_{3}t^{3}+....
\notag \\
\theta & =\theta_{0}+\theta_{1}t+\theta_{2}t^{2}+\theta_{3}t^{3}+.....
\notag \\
v & =v_{0}+v_{1}+v_{2}t+v_{2}t^{2}+.....  \label{3.17} \\
U^{\ast} & =u_{0}+u_{1}t+u_{2}t^{2}+u_{3}t^{3}+....  \notag \\
U_{\varphi}^{\ast} & =\mu_{0}+\mu_{1}t+\mu_{2}t^{2}+\mu_{3}t^{3}+....  \notag
\\
U_{\theta}^{\ast} & =\eta_{0}+\eta_{1}t+\eta_{2}t^{2}+\eta_{3}t^{3}+...
\notag
\end{align}
For convenience, some of the initial coefficiens are
\begin{align}
u_{0} & =U^{\ast}(\varphi_{0},\theta_{0})\text{, \ \ \ }u_{1}=\mu_{0}%
\varphi_{1}+\eta_{0}\theta_{1}\text{, etc.}  \notag \\
f_{0} & =\sin(2\varphi_{0}),\text{ }f_{1}=2\cos(2\varphi_{0})\varphi _{1}%
\text{, etc.}  \label{coeff} \\
\text{\ }g_{0} & =\sin^{2}(\varphi_{0}),\text{ \ }g_{1}=f_{0}\varphi _{1}%
\text{, etc.}  \notag \\
v_{1} & =\frac{1}{v_{0}}[2\varphi_{1}\varphi_{2}+\sin(\varphi_{0})\cos(%
\varphi_{0})\varphi_{1}\theta_{1}^{2}+2\sin^{2}(\varphi_{0})\theta
_{1}\theta_{2}]  \notag
\end{align}
where $v_{1}$ follows from (\ref{3.2}), and we also write
\begin{align}
\sin(2\varphi) & =f_{0}+f_{1}t+f_{2}t^{2}+....  \notag \\
\sin^{2}(\varphi) & =g_{0}+g_{1}t+g_{2}t^{2}+....  \notag
\end{align}
We shall regard $\mu_{0}$, $\eta_{0}$ as intrinsic data, but they depend on
the coordinate system, of course.

Below we shall investigate dependence relations among the coefficients in (%
\ref{3.17}) such as $\rho_{i}$, $\varphi_{j}$, $\theta_{k}$ and various
other coefficients. Some of them are directly expressible in terms of the
intrinsic data and hence regarded as constants, whereas the others are the
\emph{variables}.

\begin{definition}
\label{vari}The following list of coefficients in the expansions (\ref{3.17}%
)
\begin{equation}
\rho_{0},v_{0};\rho_{1},\varphi_{1},\theta_{1};\rho_{2},\varphi_{2},%
\theta_{2}  \label{8variables}
\end{equation}
will be referred to as the variables of order $\leq2$. The variables of
order $n$ are $\rho_{n},\varphi_{n},\theta_{n}$ when $n>0$, and $%
\rho_{0},v_{0}$ are the only variable of order zero.
\end{definition}

Henceforth, assume the above moduli curve $\bar{\gamma}(t)$ is a solution of
the ODE system (\ref{3.12})-(\ref{3.13}) with $e=1$. By inserting the power
series into the equations (i)-(iv) and applying the method of undetermined
coefficients, we arrive at the following scheme of recursive relations for
the variables of increasing order $0,1,2..:$\qquad%
\begin{align}
E_{10} & :0=2\rho_{0}^{2}\rho_{2}+\rho_{0}\rho_{1}^{2}-2h\rho_{0}-u_{0}
\notag \\
E_{20} & :0=2\rho_{0}^{3}\varphi_{2}+2\rho_{0}^{2}\rho_{1}\varphi_{1}-\frac{1%
}{2}\rho_{0}^{3}f_{0}\theta_{1}^{2}-4\mu_{0}  \label{E0} \\
E_{30} &
:0=2g_{0}\rho_{0}^{3}\theta_{2}+2g_{0}\rho_{0}^{2}\rho_{1}\theta_{1}+%
\rho_{0}^{3}f_{0}\varphi_{1}\theta_{1}-4\eta_{0}  \notag \\
E_{40} & :0=\rho_{0}^{2}\rho_{1}^{2}+\frac{1}{4}\rho_{0}^{4}(%
\varphi_{1}^{2}+g_{0}\theta_{1}^{2})+\omega^{2}-2u_{0}\rho_{0}-2h\rho_{0}^{2}
\notag
\end{align}
and in general%
\begin{align}
E_{1n} & :0=(n+2)(n+1)\rho_{0}^{2}\rho_{n+2}+.....  \notag \\
E_{2n} & :0=(n+2)(n+1)\rho_{0}^{3}\varphi_{n+2}+.....  \label{En} \\
E_{3n} & :0=(n+2)(n+1)g_{0}\rho_{0}^{3}\theta_{n+2}+......  \notag
\end{align}
where the remaining terms are of less order since they involve $\rho
_{i},\varphi_{i},\theta_{i}$ for $i<n+2$. For example, the order of the
coefficients $u_{n},$ $\mu_{n},\eta_{n}$ in (\ref{En}) is $n$. The equations
$E_{4n}$ for $n>0$ will not be needed and hence omitted since they do not
lead to additional (algebraic independent) relations.

Now, let us select some independent and recursive relations from the above
ones, but first we take the basic identity (\ref{S8}) and the expression (%
\ref{3.2}) for the speed in the spherical metric, whose zero order terms
yield the two identities :
\begin{align}
E_{0} & :\rho_{0}^{3}v_{0}^{2}=4\mathfrak{S}_{0}  \label{7a} \\
E_{0}^{\prime} & :v_{0}=\sqrt{\varphi_{1}^{2}+g_{0}\theta_{1}^{2}}
\label{7b}
\end{align}
We shall use the symbols $J_{1},J_{2}$ etc. to denote various expressions
which are of intrinsic type. By using (\ref{7a}) the identities $E_{10}$ and
$E_{40}$ can be restated as
\begin{align}
E_{1} & :\rho_{0}(\rho_{1}^{2}-2h)+\frac{\omega^{2}}{\rho_{0}}=J_{1},\text{
\ \ \ }J_{1}=2u_{0}-\mathfrak{S}_{0}  \label{list1-4} \\
E_{4} & :\rho_{0}^{2}\rho_{2}-\frac{\omega^{2}}{2\rho_{0}}=J_{4},\text{ \ \
\ \ \ \ \ \ \ \ \ \ \ \ }J_{4}=\frac{1}{2}(-u_{0}+\mathfrak{S}_{0})  \notag
\end{align}
Next, the direction $\psi_{0}$ of $\gamma^{\ast}$ at the point $(\varphi
_{0},\theta_{0})$ is intrinsic; it is also conveniently represented by the
unit tangent vector

\begin{equation*}
\mathbf{\tau}^{\ast}=\frac{1}{v_{0}}(\varphi_{1}\frac{\partial}{\partial
\varphi}+\theta_{1}\frac{\partial}{\partial\theta})=J_{\varphi}\frac{%
\partial }{\partial\varphi}+J_{\theta}\frac{\partial}{\partial\theta}
\end{equation*}
The coefficients $J_{\varphi}$, $J_{\theta}$ are intrinsic functions,
depending on the coordinate system, and they are related by the identity
\begin{equation}
J_{\varphi}^{2}+g_{0}J_{\theta}^{2}=1  \label{J-identity}
\end{equation}
Therefore, we adjoin to our list (\ref{list1-4}) the two identities
\begin{align}
E_{2} & :\varphi_{1}=J_{\varphi}v_{0}\text{ \ }  \label{list2-3} \\
E_{3} & :\theta_{1}=J_{\theta}v_{0}\text{\ }  \notag
\end{align}

Still, we have not used all zero order relations, namely $E_{20}$ and $%
E_{30} $, and now we state them as%
\begin{align}
E_{5} & :\rho_{0}^{3}\varphi_{2}+\rho_{0}^{2}\rho_{1}\varphi_{1}=J_{5},\text{
\ \ \ }J_{5}=2\mu_{0}+f_{0}J_{\theta}^{2}\mathfrak{S}_{0}  \label{list5-6} \\
E_{6} & :\rho_{0}^{3}\theta_{2}+\rho_{0}^{2}\rho_{1}\theta_{1}=J_{6},\text{
\ \ \ }J_{6}=\frac{2\eta_{0}}{g_{0}}-2\frac{f_{0}}{g_{0}}J_{\varphi}J_{%
\theta }\mathfrak{S}_{0}  \notag
\end{align}
By continuing this way, we obtain for each $n>0$ three new relations with
leading terms as indicated%
\begin{align}
E_{3n+1} & :0=\rho_{0}^{2}\rho_{n+2}+.....  \notag \\
E_{3n+2} & :0=\rho_{0}^{3}\varphi_{n+2}+.....  \label{list7-9} \\
E_{3n+3} & :0=\rho_{0}^{3}\theta_{n+2}+......  \notag
\end{align}
where the triple $\rho_{n+2},\varphi_{n+2},\theta_{n+2}$ are the variables
of highest order $n+2$. \

\begin{claim}
\label{possible}It is possible to solve the above recursive relations for
the variables (\ref{8variables}) completely in terms of the intrinsic local
geometric data in the shape space.
\end{claim}

We proceed as follows. At this point, we observe first that there are
altogether 3n+8 variables
\begin{equation*}
\rho_{0},v_{0};\rho_{1},\varphi_{1},\theta_{1};\rho_{2},\varphi_{2},\theta
_{2};...;\rho_{n+2},\varphi_{n+2},\theta_{n+2};
\end{equation*}
involved in 3n+8 recursive relations, and the first eight involve only the
variables up to order 2. However, $E_{0}^{\prime},E_{2}$ and $E_{3}$ are
obviously algebraic dependent due to the identity (\ref{J-identity}), so we
shall search for one more independent relation among the variables of order $%
\leq2$. We expect such a relation to involve local intrinsic quantities of
order (at least) 3, so a natural approach is to differentiate the basic
identity (\ref{S8}) involving the function $\mathfrak{S}$. Then, evaluation
of the resulting identity at $t=t_{0}$ yields
\begin{equation}
3\frac{\rho_{1}}{\rho_{0}v_{0}}+2\frac{v_{1}}{v_{0}^{2}}=J_{7},\text{ \ \ \
\ \ }J_{7}=\frac{\mathfrak{S}_{1}}{\mathfrak{S}_{0}}  \label{E1prime}
\end{equation}
Using the expression in (\ref{coeff}) for $v_{1}$ we can restate the above
identity as
\begin{equation}
3\frac{\rho_{1}}{\rho_{0}v_{0}}+\frac{4}{v_{0}^{3}}\left[ \varphi_{1}%
\varphi_{2}+\frac{1}{4}f_{0}\varphi_{1}\theta_{1}^{2}+g_{0}\theta_{1}%
\theta_{2}\right] =J_{7}  \label{10}
\end{equation}
By simple calculation and substitution using some of the previous relations $%
E_{i}$,
\begin{align*}
& \rho_{0}^{3}\left[ \varphi_{1}\varphi_{2}+\frac{1}{4}f_{0}\varphi
_{1}\theta_{1}^{2}+g_{0}\theta_{1}\theta_{2}\right] \\
& =\varphi_{1}(J_{5}-\rho_{0}^{2}\rho_{1}\varphi_{1})+\varphi_{1}(\frac{1}{4}%
f_{0}J_{\theta}^{2}\rho_{0}^{3}v_{0}^{2})+\theta_{1}(g_{0}J_{6}-g_{0}%
\rho_{0}^{2}\rho_{1}\theta_{1}) \\
& =-\frac{\rho_{1}}{\rho_{0}}\rho_{0}^{3}(\varphi_{1}^{2}+g_{0}%
\theta_{1}^{2})+\varphi_{1}(J_{5}+f_{0}J_{\theta}^{2}\mathfrak{S}%
_{0})+\theta_{1}g_{0}J_{6} \\
& =-4\frac{\rho_{1}}{\rho_{0}}\mathfrak{S}_{0}+v_{0}\left[
J_{\varphi}J_{5}+f_{0}J_{\varphi}J_{\theta}^{2}\mathfrak{S}%
_{0}+g_{0}J_{\theta}J_{6}\right]
\end{align*}
and by substitution into (\ref{10}), using the identity $%
\rho_{0}^{3}v_{0}^{2}$ $=4\mathfrak{S}_{0}$ and the expressions for $%
J_{5},J_{6}$ in (\ref{list5-6}), this leads to our new identity
\begin{equation}
E_{1}^{\prime}:\frac{\rho_{1}}{\rho_{0}v_{0}}=J_{8},\ \text{ \ \ \ \ \ \ }%
J_{8}=2\mathfrak{S}_{0}^{-1}(J_{\varphi}\mu_{0}+J_{\theta}\eta_{0})-J_{7}=%
\text{\ }\frac{1}{\mathfrak{S}_{0}}(2\bar{u}_{1}-\mathfrak{S}_{1})\text{\ \
\ }  \label{11}
\end{equation}
where $\bar{u}_{1}$ is the tangential derivative $U_{\tau}^{\ast}$ of $%
U^{\ast}$ at $P_{0},$ cf. (\ref{3.16}).

From the system of algebraic equations%
\begin{equation*}
E_{0},E_{0}^{\prime},E_{1},E_{1}^{\prime},E_{2},E_{3},E_{4},.....
\end{equation*}
we can now solve recursively and thus determine the variables%
\begin{equation*}
\rho_{0},v_{0,}\rho_{1},\varphi_{1},\theta_{1},\rho_{2},\varphi_{2},\theta
_{2},.....
\end{equation*}
successively in terms of the intrinsic data. In fact, this is obvious from
the structure of the equations, once we have determined $\rho_{0},v_{0},%
\rho_{1}$, namely using the three equations $E_{0},E_{1},E_{1}^{\prime}$:
\begin{equation}
\rho_{0}^{3}v_{0}^{2}=4\mathfrak{S}_{0},\text{ \ }\rho_{0}(\rho_{1}^{2}-2h)+%
\frac{\omega^{2}}{\rho_{0}}=J_{1},\text{ \ }\frac{\rho_{1}}{\rho _{0}v_{0}}%
=J_{8}  \label{3rel}
\end{equation}

It follows that $\rho_{0}$ is characterized as a positive root of the
following polynomial of order $\leq2:$
\begin{equation}
2h\rho_{0}^{2}-(4J_{8}^{2}\mathfrak{S}_{0}-J_{1})\rho_{0}-\omega^{2}=0
\label{2order}
\end{equation}
The case $\omega=0$ is discussed in \cite{HS-1}, Section 4.1, and we recall
the three cases $h=0,h>0,h<0$ are characterized by the sign of $4J_{8}^{2}%
\mathfrak{S}_{0}-J_{1}$, namely the cases are
\begin{equation}
u_{0}=\frac{1}{2}(4J_{8}^{2}+1)\mathfrak{S}_{0},\text{ \ }u_{0}<\frac{1}{2}%
(4J_{8}^{2}+1)\mathfrak{S}_{0},\text{ \ \ }u_{0}>\frac{1}{2}(4J_{8}^{2}+1)%
\mathfrak{S}_{0}  \label{3eq}
\end{equation}
Now, assume $\omega>0$. In the case of $h=0$ we have clearly%
\begin{equation}
\rho_{0}=\frac{\omega^{2}}{J_{1}-4J_{8}^{2}\mathfrak{S}_{0}}\text{ \ , \ \ \
}u_{0}>\frac{1}{2}(4J_{8}^{2}+1)\mathfrak{S}_{0}  \label{root1}
\end{equation}
For $h>0$ the solution must be
\begin{equation}
\rho_{0}=\frac{1}{4h}\left( (4J_{8}^{2}\mathfrak{S}_{0}-J_{1})+\sqrt {%
(4J_{8}^{2}\mathfrak{S}_{0}-J_{1})^{2}+8h\omega^{2}}\right)  \label{root2}
\end{equation}
but as a further characterization we cannot rule out any of the three types
of constraints (\ref{3eq}).

In the third case, $h<0,$ the two roots of equation (\ref{2order}) would be
negative if $4J_{8}^{2}\mathfrak{S}_{0}-J_{1}>0$, so the inequality "$\leq$
" must hold and consequenly
\begin{align}
\rho_{0} & =\frac{1}{4h}\left( (4J_{8}^{2}\mathfrak{S}_{0}-J_{1})-\sqrt{%
(4J_{8}^{2}\mathfrak{S}_{0}-J_{1})^{2}+8h\omega^{2}}\right)  \label{root3} \\
\text{\ } & u_{0}-\frac{1}{2}(4J_{8}^{2}+1)\mathfrak{S}_{0}\geq\omega \sqrt{%
2\left\vert h\right\vert }>0  \notag
\end{align}
The other choice of root in the formula for $\rho_{0}$ is ruled out by
demanding continuous dependence on the parameters, e.g. $\omega\rightarrow0$
should not imply $\rho_{0}\rightarrow0$.

Finally, with the above value for $\rho_{0}$, the system (\ref{3eq}) yields
the following intrinsic formulae for the two variables $v_{0},\rho_{1}$,
namely \
\begin{equation}
v_{0}=2\sqrt{\frac{\mathfrak{S}_{0}}{\rho_{0}^{3}}},\ \ \ \rho_{1}=2J_{8}%
\sqrt{\frac{\mathfrak{S}_{0}}{\rho_{0}}}  \label{2vari}
\end{equation}

\subsection{Summary and final proofs}

To complete the proofs of Theorem \ref{A} and Theorem \ref{B}, let us start
with a curve $\bar{\gamma}(t)=(\rho(t),\varphi(t),\theta(t))$ in $\bar{M}$
which is the moduli curve of a trajectory $\gamma(t)$ of a planary 3-body
motion. Then the curve $\gamma(t)$ in $M$ is uniquely determined, up to a
global congruence, by the curve $\bar{\gamma}(t)$ and the size $\left\vert
\mathbf{\Omega}\right\vert $ of the angular momentum vector. We refer to (%
\cite{HS-0}, Theorem B) for the purely kinematic result concerning the
general lifting of curves $\bar{\gamma}(t)$ in $\bar{M}$ to curves in $M$.
On the other hand, by the formula (\ref{S8}), the size function $\rho(t)$ is
already determined by the shape curve $\gamma^{\ast}(t)=(\varphi
(t),\theta(t))$ ($\gamma^{\ast}$ assumed to be non-exceptional), and this
proves Theorem \ref{A}.

In Section 3.2 it is demonstrated that the power\ series expansion of $%
\gamma^{\ast}(t)$ is essentially determined by quantities which depend only
on the geometric (i.e. unparametrized) shape curve\ $\gamma^{\ast}$ and the
relative geometry between\ $\gamma^{\ast}$ and the gradient vector field $%
\nabla U^{\ast}$. By "essential" we mean that the same shape curve can only
be reparametrized in the trivial way, namely by an affine transformation of
time, in order to remain the time parametrized shape curve of a (planary)
3-body motion. In view of Remark 1.3 this completes the proof of Theorem \ref%
{B}.\ \ \ \ \ \ \ \ \ \ \

\end{document}